\documentclass[nofootinbib,aps,floatfix,twocolumn,prl,preprintnumbers,superscriptaddress]{revtex4-2}
\pdfoutput=1
\usepackage{graphicx}
\usepackage{bm}
\usepackage{mathrsfs}
\usepackage{epstopdf}
\usepackage{url}
\usepackage{footnote}
\usepackage{textcomp}
\usepackage{dsfont}
\usepackage{ulem}
\usepackage{hyperref}
\usepackage{enumerate}   
\usepackage{amsmath, amssymb, amscd, amsthm, amsfonts}
\usepackage{hyperref}
\usepackage[dvipsnames]{xcolor}
\usepackage{subcaption}
\makeatletter
\newcommand*{\rom}[1]{\expandafter\@slowromancap\romannumeral #1@}
\makeatother

\begin{document}
	
	\title{The four-dimensional null energy condition as a swampland conjecture}
	
	\author{Heliudson Bernardo}
	\email{heliudson@hep.physics.mcgill.ca}
	\affiliation{Department of Physics, McGill University, Montr\'eal, QC H3A 2T8, Canada}
	
	\author{Suddhasattwa Brahma}
	\email{suddhasattwa.brahma@gmail.com}
	\affiliation{Department of Physics, McGill University, Montr\'eal, QC H3A 2T8, Canada}
	
	\author{Keshav Dasgupta}
	\email{keshav@hep.physics.mcgill.ca}
	\affiliation{Department of Physics, McGill University, Montr\'eal, QC H3A 2T8, Canada}
	
	\author{Mir Mehedi Faruk}
	\email{mir.faruk@mail.mcgill.ca}
	\affiliation{Department of Physics, McGill University, Montr\'eal, QC H3A 2T8, Canada}
	
	\author{Radu Tatar}
	\email{Radu.Tatar@Liverpool.ac.uk}
	\affiliation{Department of Mathematical Sciences, University of Liverpool, Liverpool, L69 7ZL, United Kingdom}

\begin{abstract}
	\noindent We analyze four-dimensional Friedmann-Lema\^{i}tre-Robertson-Walker cosmologies in type IIB, arising from a M-theory dual, and find that the null energy condition (NEC) has to be obeyed by them (except for the negatively curved case) in order for the M-theory action to have a Wilsonian effective description. However, this does not imply that the M-theory metric has to obey the $11d$ NEC. Thus, we propose a new \textit{swampland conjecture} --  the $4d$ NEC is a consistency condition for any theory to have a completion within M-theory -- with an explicit derivation of it for cosmological backgrounds from a top-down perspective. We briefly discuss the cosmological consequences of such a condition derived from M-theory.
\end{abstract}

\maketitle

\noindent  {\bf Introduction:} Energy conditions are considered important for constraining physically-viable solutions of Einstein's equations. In particular, the null energy condition (NEC) plays a crucial role in cosmology and is a key ingredient for proving the Hawking-Penrose singularity theorems \cite{Hawking-Pensrose}. The NEC implies that the matter stress-energy tensor should satisfy
\begin{eqnarray}
	T_{\mu\nu} l^\mu l^\nu \geq 0\,,
\end{eqnarray}
for a light-like vector $l^\mu$. On assuming general relativity (GR), one finds the Ricci convergence condition $R_{\mu\nu} l^\mu l^\nu \geq 0$. Although the NEC seems to be a reasonable restriction, there is no compelling derivation of it from fundamental theory \cite{Curiel} (see, however, \cite{NEC1, NEC2, NEC3, NEC4,NEC5, NEC6} for some preliminary attempts in this direction). On the other hand, there are many effective field theories (EFTs) which can violate the NEC and \`a priori, there does not seem to be a good reason to banish them \cite{Rubakov, Arkani-Hamed, Galileon}. Since these theories lead to compelling cosmological model-building with interesting physical implication (\textit{e.g.} see \cite{NEC_cosmo1,NEC_Cosmo2, NEC_Cosmo3, NEC_Cosmo4,NEC_Cosmo5}), finding a foundational origin of the NEC, as a \textit{physically necessary} condition, would have profound consequences. For instance, bouncing cosmologies --  solutions which posit that an expanding universe is created from a previously collapsing one -- present an alternative to the standard cosmological paradigm \cite{Bounce_review} and typically require NEC violation \cite{NEC_bounce1, NEC_bounce2, NEC_bounce3, NEC_bounce4}. The goal of this letter is to show a surprising link between the $4d$ NEC and a general consistency condition emanating from M-theory, thereby ruling out such bouncing solutions in string theory. 

Although supergravity theories, which are low-energy limits of string theory, have stress-energy tensors which do obey some of the energy conditions, there is no general expectation that energy conditions have to be satisfied in string theory due to some inherent fundamental reason. In fact, there exists all kinds of higher curvature terms, quantum corrections and other stringy objects (such as orientifolds and branes) which indicate that energy conditions can easily be violated in string theory. Generally speaking, the role of energy conditions in string theory is also quite well-known. The strong energy condition has been used to derive no-go theorems for having $4d$ de Sitter space descending from low-energy supergravity actions (in the absence of quantum corrections) in higher dimensions \cite{Gibbons:1984kp,Maldacena-Nunez}. On the other hand, some of the stringy effects mentioned above \cite{GKP}, as well as time-dependent internal dimensions, allow one to bypass this and find accelerating cosmologies (see, for instance, \cite{KKLT, KKLMMT} or \cite{dS_review} a review). What is clear, however, is the key role energy conditions play in understanding the types of cosmological solutions which are allowed in string theory (\cite{Russo-Townsend} presents a recent overview). Keeping this in mind, our main objective is to arrive at a remarkable derivation of {\it the four-dimensional} NEC starting from M-theory. We will derive a condition which comes from requiring that M-theory has a well-defined EFT description and show that it has \textit{precisely} the same form as the NEC in $4d$. Crucially, this \textit{will not imply} that the NEC has to be satisfied in full (higher-dimensional) M-theory but is only a \textit{consequence for the external spacetime}.


Recently, there has been a considerable effort in identifying universal features of quantum gravity which would help us in demarcating consistent EFTs in $4d$ that have a UV-completion, namely the \textit{swampland program} \cite{Vafa_review, Palti_review, Grana_review}. Our work takes a significant stride in this direction by identifying a top-down condition from M-theory which requires that any consistent  $4d$ EFT containing gravity \textit{must satisfy the NEC} in order to find an embedding in M-theory. In this way, we find a compelling reason to elevate the status of the $4d$ NEC to a \textit{swampland conjecture} -- a necessary condition that any $4d$ EFT has to satisfy in order to have a UV-completion within string theory. This shall have a lot of striking consequences for many cosmological models in $4d$. In particular, an important implication of this is that since it is well-known that violating the NEC is a necessary condition for the existence of bounces in flat Friedmann-Lemaitre-Robertson-Walker (FLRW) cosmologies, we show that such bounces cannot arise in theories descending from M-theory with a well-defined Wilsonian effective action, supporting previous similar claims from other considerations \cite{Engelhardt:2015gla, Horowitz:2003he, Craps:2007ch, Hertog:2005hu}.

Let us sketch our main result which can be understood as follows. One starts with an $11d$ M-theory metric which allows for a $4d$ FLRW spacetime and a time-dependent (warped) internal spacetime (in the dual type IIB side) and includes all  types of (time-dependent) fluxes and local and non-local quantum corrections (including higher curvature terms) that are needed to support such a spacetime \cite{coherent1, desitter2, desitter3}. We shall then derive a necessary condition for all these quantum terms to have a hierarchy, so as to have a well-defined Wilsonian effective action, which will impose a constraint on the allowed form of $a(t)$ for the external dimensions\footnote{In fact, time-dependence and Wilsonian effective action require the M-theory configuration to be realized as a Glauber-Sudarshan state and the corresponding fluctuations as an Agarwal-Tara state \cite{coherent1, coherent2} (see also \cite{coherent3, coherent4}). We will, however, not deal with these subtleties here.}. Naturally, allowing such flux sources, stringy extended objects and quantum corrections imply that the higher-dimensional metric do not obey simple two-derivative Einstein's equations. Nevertheless, one can still put all the corrections and sources to the right-hand-side of Einstein's equations and work with an \textit{effective $10d$ stress-energy tensor} which supports such a metric. What we find, quite remarkably, is that the condition required for the quantum terms to maintain their hierarchy is exactly the same as the NEC for the $4d$ external metric in the dual IIB side. Thus, as long as one has a well-defined EFT description for the fluxes and quantum terms included in the M-theory action, the $4d$ FLRW metric will \textit{automatically obey the NEC}. What is more is that the  $4d$  NEC \textit{does not imply} that the higher-dimensional metric obeys the $10d$ NEC, the latter condition not expected to arise from string theory.


\noindent {\bf The NEC from M-theory:} Let us come straight to the most novel part of our argument. On the M-theory side, let us take a metric ansatz of the form\footnote{Throughout, $\eta$ and $t$ denotes conformal and cosmic time, respectively.}:
{\small
	\begin{eqnarray}\label{M-theory_metric}
		ds^2&=&{e^{2{A}(y,\eta)}\over f^{1/3}(r, \theta)}\left(-d\eta^2+ g_{ij} dx^i dx^j\right) 
		+ {e^{2{B}(y,\eta)}\over f^{1/3}(r, \theta)}
		\tilde{g}_{mn}dy^m dy^n\nonumber\\ 	&+&
		e^{2{C}(y,\eta)} ~f^{2/3}(r, \theta)\left({d\phi^2\over g_b^2}+dx_{11}^2\right)\,,
\end{eqnarray}}
with $g_b$ being the type IIB string coupling (which is kept at the constant coupling point in F-theory), $(m,n=4,\ldots,9)$  and $(i,j= 1,2)$ and where
\begin{eqnarray}
	f(r,\theta) &=& \dfrac{1}{r^2 \sin^2\theta}\,,\,\,\,g_{11}= \dfrac{1}{1 - k r^2}\,,\,\, 	g_{22}= r^2\,.
\end{eqnarray}
Although unfamiliar, this form of the metric in M-theory simply assumes a general FLRW metric for $4$ external dimensions in the dual IIB side, for an $11d$ space which has the topology of: $\mathcal{M}_{11} = \mathbb{R}^{2,1} \times \mathcal{M}_6 \times \mathbb{T}^2/\mathcal{G}$, where $\tilde{g}_{mn} (y,\eta)$ is the unwarped metric of the $6d$ base and $\mathcal{G}$ is the isometry group. Although, as we show below and as alluded to above, there is a type IIB metric corresponding to \eqref{M-theory_metric}, we begin with this uplifted metric as it shall help us in identifying the time-dependence of the IIA string coupling which will be useful for organizing the time-dependence of all the quantum corrections and the flux 
components\footnote{There is another deeper reason for using M-theory uplift: the IIB configuration is at a constant coupling point, and $g_b = 1$ with vanishing axio-dilaton. This means it is at strong coupling (where S-duality doesn't help). M-theory uplift is the only way to allow for a controlled laboratory for the IIB computations.}. The warping factor ($\mathrm{H}(y)$) is contained in the expressions: 
\begin{eqnarray}
	&&e^{2A}= g_b^{-2/3} a(\eta)^{\frac{8}{3}}\mathrm{H}(y)^{-\frac{8}{3}}\\
	&&e^{2B}= g_b^{-2/3} a(\eta)^{\frac{2}{3}}\mathrm{H}(y)^{\frac{4}{3}}\\
	&&e^{2C}= g_b^{4/3} a(\eta)^{-\frac{4}{3}}\mathrm{H}(y)^{\frac{4}{3}}\,,
\end{eqnarray}
where $a(\eta)$ is the usual scale factor for the $4d$ cosmological metric. Dimensionally reducing the $x_{11}$ direction, we get
\begin{eqnarray}
	& &ds^2=e^{2{A}(y,\eta)+C(y,\eta)}(-d\eta^2+g_{ij}dx^idx^j)\\
	& & + e^{2{B}(y,\eta)+C(y,\eta)}	\tilde{g}_{mn}(y,\eta)dy^m dy^n + e^{3{C}(y,\eta)}f(r,\theta) 
	~{d\phi^2\over g_b^2}\,,\nonumber
\end{eqnarray}
with time-dependent type IIA coupling $g_s \equiv e^{3C/2} f^{1/2}$. In fact, we will use $g_s$ to represent the temporal behavior in the M-theory side.  Finally, T-dualizing the $\phi$ direction, we get a type IIB metric of the form\footnote{Note that $e^{2A+C}=g_b^2\, e^{-3C}=\frac{a^2(\eta)}{\mathrm{H}^2(y)}$ and $e^{2B +C} =\mathrm{H}^2(y)$.},
\begin{eqnarray}\label{IIB_metric1}
	ds^2&=&\frac{a^{2}(\eta)}{\mathrm{H}^2(y)}\left(-d\eta^2+g_{ij}dx^idx^j\right.\\
	& & \left.+ r^2\sin^2 \theta d\phi^2\right) + \mathrm{H}^2(y)\,\tilde{g}_{mn}(y, \eta)dy^m dy^n\,.\nonumber
\end{eqnarray}
As is clear from the discussion above, $y$ collectively denotes the internal spatial directions for us. Interestingly, as shown in \cite{desitter2, desitter3}, we need to allow for time-dependent fluxes for supporting such a configuration which results in a time-dependent $\tilde{g}_{mn}$.  However, we shall still require that the $4d$ Newton’s constant $G_N$ remains fixed. One can further split up the internal $6d$ manifold $\mathcal{M}_6 = \mathcal{M}_4 \times \mathcal{M}_2$, and separate out the time-dependence of it, to get
{\small
	\begin{eqnarray}\label{IIB_metric}
		ds^2 = \frac{1}{\mathrm{H}^2(y)} ds^2_{\text{FLRW}}  + \mathrm{H}^2(y)\left[F_1(\eta)ds^2_{{\cal M}_2} 
		+ F_2(\eta) ds^2_{{\cal M}_4}\right],
\end{eqnarray}}
where the unwarped metric corresponding to the internal metrics $ds^2_{{\cal M}_2}$ and $ds^2_{{\cal M}_4}$ are now time-independent. In this more familiar form, the external spacetime can be clearly seen to be a FLRW cosmology and the condition to have $G_N$ constant implies that we additionally require $F_1 F_2^2=1$, and both $F_{1,2}(\eta)\rightarrow 1$ as $g_s \rightarrow 0$. Note that the $(2, 4)$ splitting of the internal manifold, while convenient, is not essential. One could have other splittings like $(1, 5), (3, 3)$ or even 
$(a_1, a_2 , ..)$ with $a_1 + a_2 + ... = 6$ as long as the internal six-volume remains time independent and,
in the limit $g_s \to 0$, remains non-singular.

Assuming the scale factor to be of the form $a(\eta) \sim \Lambda^{n/2} \eta^n$, the type IIA coupling takes the form:
\begin{eqnarray}
	g_s = \dfrac{g_b \mathrm{H}(y)}{\left(\Lambda \eta^2\right)^{n/2}\, r\sin\theta}\,.
\end{eqnarray}
Note that the late time regime is weakly-coupled: $g_s \to 0$. The important new condition required for having a well-defined hierarchy to the higher curvature and quantum terms is that time-derivatives of $g_s$ should always be given in terms of non-negative powers of $g_s$, \textit{i.e.}
\begin{eqnarray}\label{Cond}
	\dfrac{d g_s}{d\eta} \propto g_s^{(1+1/n)\; \geq 0}\;\;\;\Rightarrow\;\; \frac{1}{n}\geq -1\,. 
\end{eqnarray}
This is the crucial condition for us, the detailed derivation of which from the M-theory side can be seen in Sec-3 of \cite{coherent_longer} (and a brief outline in the Supplementary Materials \cite{SuppMat}). The main argument behind this condition is that for the dominant $g_s$ scaling of the various terms to be positive, so as to maintain a hierarchy between the different quantum and higher curvature terms, implies that the time-derivative of $g_s$ must be a positive power of $g_s$. A heuristic way to understand this condition \eqref{Cond} is that, for our analysis, we not only require that the type IIA coupling remains small so that we are in the weak-coupling limit, but also that it remains small for the regime of validity of the solution. Therefore, \eqref{Cond} ensures that both $g_s$ and its time-derivative remains small in this regime.


Having derived the key condition \eqref{Cond}, it is easy to show that this is the NEC for a $4d$ \textit{flat} ($k=0$) FLRW cosmology in disguise. We shall consider the open and closed case ($k = \pm 1$) later on. For a perfect fluid in $4d$, the NEC condition is given by $\rho+ p \geq 0$, where $\rho$ and $p$ are the energy and pressure densities, respectively. Assuming Einstein's equations (or, in other words, considering an effective stress-energy tensor) for a flat, FLRW spacetime, it is easy to show that the NEC implies that $\dot{H} \leq 0$ where we denote the Hubble parameter as $H(t)= \dot{a}/a$, written in terms of cosmic time. On assuming a power-law ansatz, $a(t) \propto t^\gamma$, the NEC implies $\gamma \geq 0$. Converting to conformal time, as was done for the metric \eqref{IIB_metric1} above,  \textit{i.e.} $a(\eta) \propto \eta^{\gamma/\left(1-\gamma\right)} =: \eta^n$, the NEC for $4d$ $k=0$ FLRW metric takes the form $\frac{1}{n} \geq -1$, which is exactly the same as \eqref{Cond}. 


\noindent {\bf No NEC for IIB metric:} Let us go back to \eqref{IIB_metric} and calculate the Riemann and the Ricci compenents for this metric, to demonstrate that
\begin{eqnarray}\label{rienec}
	R_{00}^{\rm (10\,d)} + a^{-2}R_{11}^{\rm (10\,d)} &=& -2\dot{H}-3\frac{\dot{F}_2^2}{F_2^2}\,,
\end{eqnarray}
where we have assumed $F_1 F^2_2=1$, as required. (Other splittings of the internal six-manifold introduce different positive powers of $F_i$ in \eqref{rienec}.) If the IIB metric were to obey the NEC (in $10d$), then we would get the condition
\begin{eqnarray}
	-2\dot{H}-3\frac{\dot{F}_2^2}{F_2^2} \geq 0\,.
\end{eqnarray}
However, recall that \eqref{Cond} simply implies that $ -\dot{H} \geq 0$, and therefore we cannot comment whether \eqref{IIB_metric} obeys the NEC or not. Note that this conclusion is only dependent on our requirement that the $4$d $G_N$ remains time-independent and does not depend on the details of the splitting of the internal manifold. This is a very intriguing finding and let us comment on its physical implication. Requiring that there exists a hierarchy in the various flux, curvature and quantum terms included in the M-theory action $-$ as is necessary to support a metric of the form \eqref{M-theory_metric} $-$ implies that the external $4d$ metric has to obey the NEC. But \textit{this does not imply} that the higher-dimensional metric also has to obey the NEC! Physically, this is indeed what one could have expected. Since we are allowing all sorts of higher curvature and (local and non-local) quantum corrections, along with time-dependent G-flux sources, our equations are very far away for the low-energy supergravity ones. Thus, there is no reason to expect that our effective stress-energy tensor for M-theory obeys any energy condition, including the NEC. Moreover, the higher-dimensional NEC would impose a geometric restriction which would never be reproduced from the lower dimensional spacetime since there are higher-dimensional null vectors which have vanishing components in some of the external directions. Also, note that for a given dimension, having a Wilsonian effective action does not imply anything like the NEC at all. In fact, there are well-known QFTs involving higher derivative terms which violate the NEC but have a consistent EFT description \cite{Arkani-Hamed, Galileon}. What we do find is that requiring that the $11d$ M-theory has all of its terms under control, in the sense of having a well-defined hierarchy of terms in the effective action, \textit{automatically leads} to imposing the NEC on the external $4d$ flat FLRW cosmology. This is why our result is exactly the same in spirit of the `swampland' conjectures -- we find that a large space of $4d$ Lagrangians (all those which violate the $4d$ NEC) cannot find a UV-completion into M-theory.


\noindent {\bf Advantages \& Assumptions:} At this point, let us emphasize the main assumptions in our derivation above. First, we assume that the external FLRW spacetime is parametrized by $a(\eta) \propto \eta^n$. This is true for perfect fluids in $4d$ with a constant equation of state of the form $p=w \rho$. Secondly, we require that the M-theory action must have hierarchy between quantum corrections of different orders and, therefore, has a well-defined Wilsonian EFT description. Although this is a rather conservative assumption, it might happen that there exists solutions for which one needs to take into account quantum corrections of all orders and no truncations to any finite order is allowed. Thirdly, although we allow for time-dependent fluxes and internal dimensions, we make sure that $G_N$ remains constant. And finally, for our explicit calculations, we have kept the type IIB dilaton to be time-independent although this is not a significant limitation and it will not be too difficult to relax this in the future.

Having said this, note that our analysis provides a powerful advantage over other approaches and it is rather general in the following sense. Our M-theory solution is not limited to leading order in $\alpha'$ or $g_s$ corrections. Indeed, we allow for all types of perturbative, non-perturbative and topological quantum corrections along with all possible higher-curvature terms. We find that time-dependent fluxes are necessary to support a configuration like \eqref{M-theory_metric}, which has a $4d$ external FLRW metric, so those have to be included as well. This is, in fact, what should make us skeptical about whether the higher-dimensional metric would obey anything like the NEC. More importantly, this implies that we do not constrain the effective stress-energy tensor for our M-theory solution to obey any energy conditions. Simply ensuring that there exists a hierarchy between the different terms allows us to derive \eqref{Cond}, which turns out to be the NEC for the $4d$ flat FLRW metric.


\noindent {\bf Cosmological implications:}  An immediate consequence of \eqref{Cond} is that cosmological bounces are ruled out for flat FLRW spacetimes if they have to descend from M-theory. This is not a statement for bounces in the context of classical gravity. Indeed, we are deriving a condition \eqref{Cond} from full M-theory which happens to match with the NEC for $k=0$ FLRW cosmology. We emphasize that we are constraining an \textit{effective stress-energy tensor} for the external $4d$ metric, the full embedding of which within M-theory \eqref{M-theory_metric} contains all types of flux sources and quantum corrections. Thus, we are not just considering the avoidance of singularity by some classical bouncing solution but rather commenting on the status of bouncing cosmologies having a UV-completion within string theory. There has been previous similar statements regarding banishing cosmological bounces from principles of holography \cite{Engelhardt:2015gla} or properties of initial or final boundary conditions \cite{Horowitz:2003he}. However, our argument comes from a much more general principle and is therefore, applicable to a much wider class of cosmological models. The fact that we have to obey the $4d$ NEC does not, of course, mean that singularity resolution is not possible in cosmological models of M-theory. One can think of a situation where the bouncing solution requires corrections to all orders and so has no effective description \cite{Alpha_prime}, or even where spacetime is emergent from more fundamental UV degrees of freedom \cite{Emergent}.

The second important conclusion of having to obey the NEC is that the Hubble parameter can neither grow today or in the early universe (superinflation) \cite{Superinflation}. More specifically, models of dark energy which violate the NEC would be ruled out immediately insofar that they can have no quantum gravity completion. A large class of dark energy models which require an equation of state $w<-1$ is immediately ruled out. This is of enormous phenomenological importance since the recent Hubble tension is seemingly alleviated by dark energy models which have a phantom component \cite{H0-1, H0-3} and our condition \eqref{Cond} would severely disfavor such Lagrangians \cite{H0-2}.

Further consequences of having the NEC as a swampland condition is that it would rule out traversible Lorentzian wormholes in $4d$ \cite{Wormhole-1, Wormhole-2} and creating a universe in a laboratory \cite{Farhi-Guth}. Moreover, all NEC violating FLRW cosmologies, such as what one gets from a large subclass of modified gravity (for instance, from Horndeski or, more generally, DHOST) theories are ruled out due to this consistency condition. Therefore, we are able to severely constrain the space of allowed cosmological models which come from a plethora of $4d$ gravitational theories, if they are to have a UV-complete description. 

Let us end our discussion of cosmological implications with an important disclaimer. It is important to emphasize that our condition \eqref{Cond} only exhibits the NEC for $4d$ and not the strong energy condition (SEC). Had we found the latter, it would have ruled accelerating solutions in cosmology for M-theory. However, as discussed in \cite{desitter2, coherent1}, we do find such solutions, including $4d$ de Sitter space since, as stressed above, we do \textit{not} require that the M-theory stress energy tensor to satisfy the $11d$ NEC. It has been independently shown that satisfying the SEC, even if the NEC is violated, allows for time-depending compactifications to 4d de Sitter space \cite{Russo-Townsend}.


\noindent {\bf Generalization to $k=\pm 1$:} Note that our condition \eqref{Cond}, which comes from M-theory, is true for any curvature \eqref{M-theory_metric}. It just so happens that this coincides with the NEC for the flat FLRW metric in $4d$. In order to see what our condition implies for the closed ($k=1$) and the open ($k=-1$) case, let us write down the NEC for a general FLRW metric. Assuming $a(t) \sim t^\gamma$, as before, we find the NEC implies
\begin{eqnarray}\label{NEC_general}
	-\dfrac{d^2}{dt^2} \left(\ln t^\gamma\right) +k\, t^{-2\gamma} \geq 0 ~~
	\Rightarrow \; \gamma + k\, t^{2\left(1-\gamma\right)} \geq 0\,.
\end{eqnarray}
This confirms our previous assertion that the NEC for $k=0$ is $\gamma \geq 0$, which written in terms of scale factor expressed in conformal time translates to \eqref{Cond}. The above equation \eqref{NEC_general} immediately tells us that given \eqref{Cond}, the NEC is also going to be automatically satisfied for closed FLRW spacetime. This is because, for the $k=1$ case, the second term in \eqref{NEC_general} is necessarily positive. Finally, for the negatively curved $k=-1$ case, \eqref{Cond} does not imply the NEC condition in this case. To summarize, we find that our new swampland conjecture would be that any EFT in $4d$ \textit{closed or flat} FLRW cosmology must satisfy the NEC in order to have a UV-completion. 

A final point to note is that this also indicates that our analysis rules out cosmological bounces \textit{only for flat FLRW spacetimes}. This is so because violating the NEC is a necessary condition for having a bounce only for flat and open FLRW metrics (and we do not get the NEC for the open case). In fact, although condition \eqref{Cond} implies the NEC for the closed case, one can have cosmic bounces for this geometry without violating the NEC. Therefore, our conclusions regarding bounces are only limited to the $k=0$ FLRW spacetime. Having said this, let us note that a generic contracting solution is unstable against anisotropies and can even become strongly inhomogeneous due to the BKL conjecture. To avoid this stability problem of bounces, a typical attractor solution is often invoked known as \textit{ekpyrosis} \cite{Ekpyrosis1, Ekpyrosis2}, which assumes a super-stiff equation of state $w \gg 1$. However, what this physically implies is that near the bounce, at the end of the contracting phase, the term proportional to the curvature $k$ is sub-leading and therefore, for ekpyrotic scenarios, our results would generically apply. In other words, although we have a condition which explicitly rules out cosmological bounces for flat FLRW spacetimes, in effect we find a very strong argument against all types of bounces since a contracting solution that is stable against anisotropies is agnostic about the curvature of spacetime anyway.


\noindent {\bf Conclusion:} In this letter, we showed that there appears a remarkable connection between the requirement of having a well-defined Wilsonian EFT for M-theory (with a time-dependent compactification) and the NEC in four-dimensions. This led us to conclude that we can rule out bounces in M-theory, at least for flat FLRW cosmologies.  Since we explicitly \textit{derive} the NEC only for $k=0$ and $k=1$ FLRW spacetimes starting from M-theory, in the spirit of the swampland, we conjecture that: \textit{Any $4$d consistent theory of gravity must obey the NEC in order to have an embedding in M-theory}. We emphasize that our conjecture, for the specific $4$d FLRW backgrounds mentioned above, is actually derived from M-theory without any bottom-up considerations.

\bigskip

\section*{Supplementary Materials}
In our letter, we have shown how a consistency condition from M-theory looks exactly like the $4$d NEC for a flat, FLRW spacetime. A detailed derivation of this condition can be found in \cite{coherent_longer}. Here, we give a brief outline of this derivation by considering the hierarchy of quantum terms in the Wilsonian effective action for M-theory. 

Let us first recall our key condition as:
\begin{eqnarray}\label{Cond}
	\dfrac{d g_s}{d\eta} \propto g_s^{(1+1/n)\;\geq 0}\;\;\;\Rightarrow\;\; \frac{1}{n}\geq -1\,. 
\end{eqnarray}
We will use a Wilsonian effective action, meaning that the perturbative quantum terms may be succinctly presented as the following action
\begin{widetext}
	\begin{eqnarray}\label{quantum}
		{\bf S}_{quantum} = \sum_{l, n}{\rm M}_p^{11 - \sigma_{nl}} c_{nl}\int d^{11}x\sqrt{-{\bf g}_{11}}
		\left[{\bf g}^{-1}\right] \prod_{i = 0}^4 \left[\partial\right]^{n_i} 
		\prod_{{\rm k} = 1}^{60} \left({\bf R}_{\rm A_k B_k C_k D_k}\right)^{l_{\rm k}} \prod_{{\rm r} = 61}^{100} 
		\left({\bf G}_{\rm A_r B_r C_r D_r}\right)^{l_{\rm r}},
	\end{eqnarray}
\end{widetext}
where $c_{nl}$ are constant coefficients (independent of ${\rm M}_p$) with $n \equiv \{n_i\}, l \equiv \{l_i\}$; 
all the bold-faced quantities are defined in terms of the $g_s$ dependent warped-metric and there are sixty possible curvature terms and forty possible G-flux components, modulo their permutations. All are raised to various powers of $(l_k, l_r)$, which are in turn further acted on by five possible derivative actions: derivatives along the temporal, spatial, internal ${\cal M}_4$ and ${\cal M}_2$ as well as toroidal ${\mathbb{T}^2\over {\cal G}}$ directions. We will also use the derivative countings to fix the ${\rm M}_p$ scalings $\sigma_{nl}$ in \eqref{quantum}. Finally, all indices are contracted appropriately by inverse metric components, denoted symbolically by $\left[{\bf g}^{-1}\right]$ in \eqref{quantum}. There are also non-perturbative and non-local contributions, but we shall ignore them here to avoid making the analysis too complicated. However, including those terms do not modify our main argument  \cite{coherent1, coherent2} regarding the hierarchy which leads to \eqref{Cond} as briefly mentioned later.

Something interesting happens with the action \eqref{quantum}: computing the Riemann curvature terms for the metric given by Eqn. (2) of the main draft, and demanding that the G-flux components satisfy the following ansatze
{\small
	\begin{eqnarray}
		{\bf G}_{\rm ABCD}({\bf x}, y, z; g_s) \equiv \sum_{k \in {\mathbb{Z}\over 2}} 
		{\cal G}^{(k)}_{\rm ABCD}({\bf x}, y, z)\left({rg_s~\sin~\theta\over g_b{\rm H}}\right)^{l_{\rm AB}^{\rm CD} +
			{2k\over 3}}
\end{eqnarray}}
where $({\bf x}, y, z) \in \left(\mathbb{R}^2, {\cal M}_4 \times {\cal M}_2, {\mathbb{T}^2\over {\cal G}}\right)$ and $l_{\rm AB}^{\rm CD}$ is the dominant scaling, one can show that the quantum pieces in \eqref{quantum}
scale as some specific powers $\theta_{nl}$ of $g_s$, {\it i.e.} as 
$\left({rg_s~\sin~\theta\over g_b{\rm H}}\right)^{\theta_{nl}}$, for given choices of $(n_i, l_k, l_r)$. Additionally, one can show \cite{desitter2, desitter3}, $\theta_{nl} =
\theta_{nl}(n_i, l_k, l_r, l_{\rm AB}^{\rm CD})$. The requirement of $g_s$ and ${\rm M}_p$ hierarchies then 
instructs us to have $\theta_{nl} > 0$ (if $\theta_{nl} \le 0$ there are unsurmountable issues with 
hierarchies \cite{evan, desitter2, desitter3}).

Let us now see where does the condition \eqref{Cond} arise in controlling the inherent hierarchies or keeping $\theta_{nl} > 0$. To this end, a simple computation of the Riemann curvature can illustrate the underlying issue.  The Riemann curvature ${\bf R}_{mnpq}({\bf x}, y; g_s)$ for $(m, n) \in {\cal M}_4 \times 
{\cal M}_2$ and assuming, for simplicity, no dependence on the toroidal direction is
\begin{widetext}
	\begin{eqnarray}\label{riemann}
		{\bf R}_{mnpq}({\bf x}, y; g_s) &\equiv& {\bf R}^{(1)}_{mnpq}({\bf x}, y)
		\left({rg_s~\sin~\theta\over g_b{\rm H}}\right)^{-{2\over 3}} + {\bf R}^{(2)}_{mnpq}({\bf x}, y)
		\left({rg_s~\sin~\theta\over g_b{\rm H}}\right)^{+{4\over 3}}\\
		&+&  
		{\bf R}^{(3)}_{mnpq}({\bf x}, y)
		\left({rg_s~\sin~\theta\over g_b{\rm H}}\right)^{-{2\over 3}}
		\left[{\partial\over \partial\eta}\left({rg_s~\sin~\theta\over g_b{\rm H}}\right)\right]^2\,,\nonumber
	\end{eqnarray}
\end{widetext}
where the first term involves derivatives acting along ${\cal M}_4 \times {\cal M}_2$, the second term involves
derivatives along $\mathbb{R}^2$, and the last term involves temporal derivative. We see that the dominant scaling of the Riemann curvature term is $g_s^{-2/3}$, which is encouraging because it 
contributes  as 
$+{2l_1\over 3}$ to the quantum scaling $\theta_{nl}$, where we assume that ${\bf R}_{mnpq}$ appears as 
$({\bf R}_{mnpq})^{l_1}$ in \eqref{quantum}\footnote{It is not too hard to see why this is the case. Consider the simple case where $l_1 = 2$. This means we have $\left({\bf R}_{mnpq}\right)^2 \equiv 
	{\bf R}_{mnpq}{\bf R}^{mnpq}$. This contributes as $-{2\over 3} \times 2 + {2\over 3} \times 4 = +{4\over 3} = 
	+{2} \times {2\over 3}$ (in the first equality $-{2\over 3}$ appears from the dominant scaling of ${\bf R}_{mnpq}$, and $+{2\over 3}$ appears from the dominant scaling of the metric component ${\bf g}_{mn}$. Since there are 2 copies of the curvature tensors and 4 copies of the metric tensors, they add up to 
	$+{4\over 3}$). For odd $l_1$, let us take the case $l_1 = 3$. This gives $\left({\bf R}_{mnpq}\right)^3 =
	{\bf R}_{mnpq}{\bf R}^{pqrs}{\bf R}_{rs}^{~~mn}$, which scales as $-{2\over 3} \times 3 + {2\over 3} \times
	6 = 3 \times {2\over 3}$. Thus
	for arbitrary even or odd $l_1$ the contribution is $+{2l_1\over 3}$.}. 
According to our analysis, {\it positive} sign is good because it preserves $g_s$ and  ${\rm M}_p$ hierarchies. However this conclusion is crucially based on the condition that ${\partial g_s\over \partial \eta}$ {\it does not change the dominant scaling in \eqref{riemann}}. 

The dominant scaling in \eqref{riemann} can change when the ${\partial g_s\over \partial \eta}$ creates {\it negative} powers of $g_s$. Such negative powers are problematic because any deviation of the dominant scaling from $-{2\over 3}$ for ${\bf R}_{mnpq}$ introduces relative {\it minus} signs in the quantum scaling $\theta_{nl}$. Such relative minus signs violate both the $g_s$ and ${\rm M}_p$ hierarchies as was shown in details in \cite{desitter2, desitter3}. Note that, since generically powers of $g_s$ jump by $g_s^{\pm{\mathbb{Z}/3}}$ (in order to satisfy the EOMs, anomaly cancellation, flux quantizations and Bianchi identities), the {\it least} negative power that  $\left({\partial g_s\over \partial \eta}\right)^2$ would contribute is $g_s^{-{2\over 3}}$, making the dominant scaling to be at least $g_s^{-{4\over 3}}$. This is already problematic because it creates time-neutral series, destroying the underlying $g_s$ and ${\rm M}_p$ hierarchies 
(see \cite{evan, desitter2, desitter3} for details). 
Thus the only way out is to invoke the condition \eqref{Cond}. 

Let us end this appendix by briefly mentioning how and where do the non-perturbative terms (as well as the non-local counter-terms) become useful. Again a simple analysis of the Riemann curvature tensors illustrates the point. Let us consider the curvature tensor ${\bf R}_{ijij}$ where $(i, j) \in {\bf R}^2$. Using similar computations as in \eqref{riemann}, we can easily infer the dominant scaling to be $\left({g_s\over {\rm HH}_o}\right)^{-{14 \over 3}}$, as long as \eqref{Cond} is satisfied. This dominant scaling further implies that the space-time Einstein tensors would scale as $g_s^{-2}$. Classically, one can show that only the flux kinetic terms for ${\bf G}_{{\rm MN}ab}$, where $({\rm M, N}) \in {\cal M}_4 \times {\cal M}_2$ and $(a, b) \in {\mathbb{T}^2\over {\cal G}}$, contribute to this order. Unfortunately however, due to a no-go theorem \cite{Gibbons:1984kp, Maldacena-Nunez}, their contributions aren't enough to solve the EOMs. The only other terms that can contribute to this order are the non-perturbative and the non-local counter-terms. As shown in \cite{desitter2, coherent1, coherent2}, their contributions precisely violate the no-go conditions by appropriately inserting quartic order curvature terms non-perturbatively via the BBS \cite{BBS} instantons. The non-local counter-terms further contribute sub-dominantly, again to the same order in $g^{-2}_s$. Interestingly, once we sum up the trans-series associated with the nonlocal counter-terms, the result is a finite  and {\it local} contribution that consistently solves the EOMs. These EOMs, in the language of the Glauber- Sudarshan state, are the Schwinger-Dyson's equations (see details in \cite{coherent1, coherent2, bernardo}).

\bigskip

\noindent {\bf Acknowledgements:}
SB is supported in part by the NSERC (funding reference \#CITA 490888-16) through a CITA National Fellowship and by a McGill Space Institute fellowship. HB, KD and MMF are supported in part by the NSERC grants.

\end{document}